\documentclass[pdftex,epjc3,twocolumn]{svjour3} 

\RequirePackage{fix-cm} 
\smartqed  
\RequirePackage{graphicx}
\usepackage{xcolor}
\usepackage{soul}
\usepackage{bm}
\usepackage{amssymb}
\usepackage{amsmath}
\usepackage{amsfonts}
\usepackage{lineno,hyperref}
\modulolinenumbers[5]

\journalname{Eur. Phys. J. C}
\begin{document}

\title{Geometrical Scaling of Direct Photons in \\ Relativistic Heavy Ion and d+Au Collisions}

\author{Vladimir Khachatryan \thanksref{e1,addr1,addr2} \and \\  Micha{\l} Prasza{\l}owicz \thanksref{e2,addr3}}
\thankstext{e1}{vladimir.khachatryan@duke.edu}
\thankstext{e2}{michal@if.uj.edu.pl}

\institute{
\label{addr1} Department of Physics, Duke University, Durham, NC 27708, USA \and
\label{addr2} Department of Physics and Astronomy, Stony Brook University,  Stony Brook, New York 11794-3800, USA \hspace{-0.8cm}\and 
\label{addr3} Institute of Theoretical Physics, Jagiellonian University,  S. {\L}ojasiewicza 11, 30-348 Krak{\'o}w, Poland\hspace{-0.6cm} 
}


\date{Received: date / Accepted: date}
\maketitle

\begin{abstract}
In this paper, we show that multiplicity spectra of direct photons in A+A and d+Au  collisions at different centrality 
classes and different energies exhibit geometrical scaling, {\em i.e.}, they depend on a specific combination of number 
of participants $N_{\rm part}$, collisions energy $W$, and transverse momentum $p_{T}$ --  called saturation scale -- rather than  
on all these three variables separately. In particular, the dependence on the geometry of collisions encoded in the dependence on 
$N_{\rm part}$ is in agreement with the expectations based on the  Color Glass Condensate theory.
\end{abstract}

\section{\label{sec:intro} Introduction}

Direct photons possibly originate from the entire space-time evolution of the matter produced in relativistic heavy ion collisions (HIC): 
from the pre-equilibrium initial stage, hot fireball of the Quark-Gluon Plasma (QGP), as well as the late hadronic phase. These photons 
are excellent probes for studying the properties and dynamics of the produced matter, and by definition do not originate from hadronic 
decays. Since they do not interact strongly while passing through the QGP environment, the information that they carry is not washed 
out by final state interactions. In p+p and p(d)+Au collisions, they are basically produced only from the initial state wave function. Therefore, 
one may expect qualitative differences in the photon transverse momentum ($p_{T}$) spectra (invariant yields) with the change of the sizes 
of colliding systems/species. Indeed, the thermal photon $p_{T}$-spectra in HIC are enhanced with respect to a p+p reference at small 
$p_{T}$, and show large anisotropy (elliptic flow)  \cite{Adare:2015lcd} --  a phenomenon  dubbed {\it thermal photon puzzle}. The photon 
puzzle is by now less surprising after the collaborations at LHC have measured elliptic flow of charged particles in p+p 
\cite{Sirunyan:2017uyl,Aaboud:2018syf,Acharya:2019vdf}, suggesting a possibility of initial state correlations not associated with 
hydrodynamical evolution. 

In this paper, we show that the direct photon $p_{T}$-spectra exhibit geometrical scaling (GS), which is a property of initial state dynamics 
associated with over-occupation of gluons  at small Bjorken $x$, and even more so in large nuclei. 
GS has been first observed in the inclusive deep inelastic electron-proton scattering (DIS)~\cite{Stasto:2000er}, where 
the reduced cross section, essentially $F_{2}(x,Q^{2})/Q^{2}$, which is in principle a function of two variables $x$ and $Q^2$, depends only 
on the scaling variable $\tau^{2} = Q^{2}/Q^{2}_{\rm sat}(x)$ for small values of $x$. In this regard, the saturation scale defined as
\begin{equation}
Q^2_{\rm sat}(x) = {\bar Q}^2 \left( x/x_0 \right)^{-\lambda} ,
\label{Qsat}
\end{equation}
is proportional to the gluon density in the proton. In the case of DIS, $x_0$ and ${\bar Q}= Q_0$ are fixed parameters of the order
of  $10^{-3}$ and 1 GeV/c, respectively. The exponent $\lambda \approx 0.33$ is a nonperturbative dynamical quantity following 
from the properties of the (non-linear) QCD evolution \cite{BK,Mueller:2002zm,Munier:2003vc}, with its numerical value fixed from the data
(for review see Ref.~\cite{McLerran:2010ub, Praszalowicz:2018vfw}). GS has also been observed in the charged particle spectra in hadronic collisions 
\cite{McLerran:2010ex,McLerran:2010wm,Praszalowicz:2015dta}. In HIC, $Q^2_{\rm sat}$ is enhanced  by the overlap of interacting 
nucleons in the colliding nuclei (see Sec.~\ref{sec:sat_sca}).

Physically, the saturation scale corresponds to the gluon density in colliding systems. However, the photons are produced from quarks 
(through $q\bar{q}$ annihilation and QCD Compton scattering), and therefore they do not probe directly the over-occupied gluonic cloud. 
To this end, one can employ the Color Glass Condensate (CGC) theory to describe the initial hadronic wave function, which evolves into an 
intermediate state called Glasma, being kind of a strongly interacting and not thermalized QGP~\cite{McLerran:2010ub}. The quarks are 
produced in the thermalization process, and -- if there are no other mass scales around -- their distribution should exhibit geometrical 
scaling\footnote{The detailed description of the photon production mechanism in Glasma is beyond the scope of this paper; we refer 
the reader to Refs.~\cite{Chiu:2012ij,McLerran:2015mda,Khachatryan:2018ori} for details.} in terms of scaling variable $\tau$:
\begin{equation}
\tau = p_{T}/Q_{\rm sat}(x), ~~{\rm where}~~x=e^{\pm y} \, p_{T}/\sqrt{s_{NN}}.
\label{eq:tau}
\end{equation}
Consequently, the direct photon $p_{T}$-spectra should scale as well. Indeed, the analysis of 
Ref.~\cite{Klein-Bosing:2014uaa} showed 
the emergence of GS in direct photon production, where the authors  assumed the functional form of the photon spectra
to be $p_{T}^{-n}$ (see also Refs.~\cite{Chiu:2012ij,McLerran:2015mda,Khachatryan:2018ori}). In this paper we extend the analysis
of  Refs.~\cite{Klein-Bosing:2014uaa,Khachatryan:2019uqn,Praszalowicz:2018vfw}
including later experimental data and employing a model-independent method to search for GS,
rather than assuming an approximate functional shape of the scaling function. Preliminary results have been already presented by
one of us in Ref.~\cite{Praszalowicz:2018vfw}.

Although the GS scenario has been postulated in Ref.~\cite{Stasto:2000er} following the idea of the saturation exemplified by the GBW 
model \cite{GolecBiernat:1998js} for deep inelastic $e-p$ scattering, it in fact arises also in the kinematical regions beyond saturation 
\cite{Iancu:2002tr,Kwiecinski:2002ep,Caola:2008xr}. Indeed, Iancu, Itacura and McLerran have shown in Ref.~\cite{Iancu:2002tr}
that the BFKL evolution preserves GS beyond saturation when starting from the boundary conditions in the saturation region. 
Kwieci{\'n}ski and Sta{\'s}to have in turn studied the DGLAP evolution in Ref.~\cite{Kwiecinski:2002ep}, showing that GS is stable for 
sufficiently large values of the parameter $\lambda$ at low $x$ in the fixed coupling case. In the running coupling case, GS has been 
found to be only approximately preserved at small $x$. These studies have been further extended by Caola and Forte in 
Ref.~\cite{Caola:2008xr}, where they have shown that an approximate GS is in fact a general property of the DGLAP evolution with 
general boundary conditions. The saturation scale emerges from a saddle point of the DGLAP kernel being a generic property of 
perturbative evolution. As such, it also generates GS for the BFKL evolution. For low $Q^2$ and small $x$, perturbative solutions 
violate GS,  therefore its emergence provides genuine evidence for parton saturation. 

Even though GS may not be a sufficient argument in favor of saturation, also in the case of hadronic collisions, it is hard to imagine that 
the scaling behaviour of direct photon $p_{T}$-spectra that we observe in this paper (following from geometrical characteristics of the 
saturation scale) can be explained without invoking the saturation mechanism.

Our paper is organized as follows. In Sec.~\ref{sec:sat_sca}, we discuss the geometry of HIC, d+Au and p+p collisions\footnote{Note
that throughout the paper we also use the expressions {\it large systems} (for HIC) and {\it small systems} (for d+Au and p+p collisions).} 
and introduce pertinent scaling variables. In Sec.~\ref{sec:GS}, we analyze available direct photon data concentrating on two aspects 
of GS: geometry and energy dependence. We summarize our findings in Sec.~\ref{sec:sumcon}.

\section{\label{sec:sat_sca} Saturation and scaling variables}

The GS physics emerges when particles are produced in a kinematical region, where the only relevant intrinsic scale is the saturation 
momentum of a radiation source. Then, by dimensional analysis, the particle multiplicity is proportional to the overlap transverse area 
$S_{T}$ and a universal function of the scaling variable (\ref{eq:tau}):
\begin{equation}
\frac{1}{S_{T}}\frac{dN_{\gamma }}{\,2\pi p_{T }\,d\eta dp_{T }}=\,F(\tau ) . 
\label{multHI}
\end{equation}%
The energy dependence of $Q_{\rm sat}$ is given by Eq.~(\ref{Qsat}). 

\begin{figure}[h]
\centering
\includegraphics[scale=0.25]{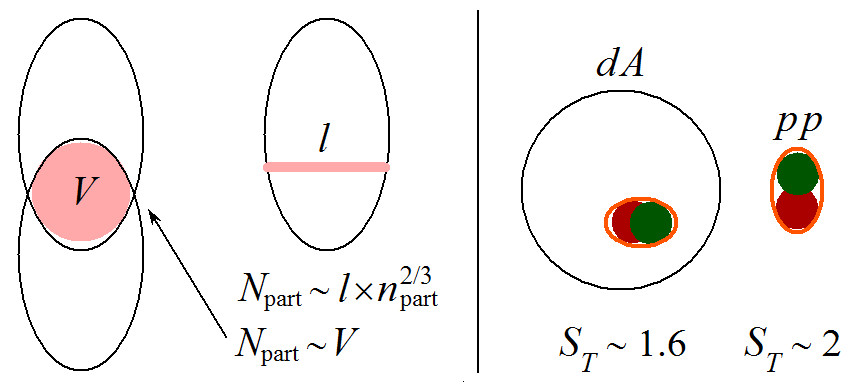}
\caption{(Color online) Overview of the collision geometry. Left panel: the side view of large and small system collisions showing 
relation of the number of participants to the collision geometry. Right panel: the front view of d+Au and p+p collisions. Active 
transverse area (shown in colors) in d+Au is entirely due to the average size of the smaller nucleus. In p+p, since collisions are 
mostly \emph{peripheral}, active transverse area is close to twice that of the proton.
}
\label{fig:geometry}
\end{figure}

As far as the value of $\bar{Q}^2$ is concerned, we have to distinguish two geometrical setups corresponding to symmetric and asymmetric 
collisions \cite{Kharzeev:2000ph,Kharzeev:2002ei}. A photon with transverse momentum $p_{T }$ in the mid-rapidity region resolves, by 
uncertainty principle, partons of the characteristic transverse size $s_{T}\sim \pi /p_{T }^{2}$. 
For symmetric A+A collisions, the interaction region approximately corresponds to a sphere of radius $R$, as shown in Fig.~\ref{fig:geometry}.
The number of all active sources of radiation, \emph{i.e.}, the number of participants $N_{\rm{part}}$, is proportional to $\pi R^{3}$.
Therefore,
\begin{equation}
R = N_{\rm{part}}^{1/3}/Q_{0},~~S_{T} = \pi N_{\rm{part}%
}^{2/3}/Q_{0}^{2} .
\label{ST}
\end{equation}%
Here $S_{T}$ is centrality-dependent transverse overlap of the colliding nuclei. When the number of active partons reaches a critical 
value $N_{\rm{cr}}$, where the whole transverse area is covered by resolved partons
\begin{equation}
N_{\rm{cr}} = \frac{S_{T}}{s_{T}} = N_{\rm{part}}^{2/3}%
\,\frac{p_{T }^{2}}{Q_{0}^{2}} ,
\label{Ncrit}
\end{equation}%
then the number of partons cannot grow any more and the system reaches the saturation, $N_{\rm{part}} \approx N_{\rm{cr}}$. 
Consequently, the average transverse momentum, $\left\langle p_{T }^{2}\right\rangle$ (related to the saturation scale) when 
this happens, calculated from (\ref{Ncrit}) is equal to%
\begin{equation}
{\bar Q}^{2} \equiv \left\langle p_{T }^{2}\right\rangle = \frac{N_{%
\rm{cr}}}{N_{\rm{part}}^{2/3}}\,Q_{0}^{2} = N_{\rm{part}}^{1/3}Q_{0}^{2} .
\end{equation}%
This means, getting back to Eqs.(\ref{Qsat}) and (\ref{eq:tau}), that for mid-rapidity photons we will have
\begin{equation}
Q_{\rm{sat}} = N_{\rm{part}}^{1/6}Q_{0}\left( \frac{p_{T }}{x_0 \sqrt{s_{NN}}}\right)
^{-\lambda /2} .
\label{Qsat_sym}
\end{equation}%
Introducing the variable $W=\sqrt{s_{{NN}}}\times x_0=\sqrt{s_{{NN}}}\times 10^{-3}$, we will further have
\begin{equation}
\tau = \frac{p_{T}}{Q_{\rm sat}} = \frac{1}{N_{\rm{part}}^{1/6}}\frac{p_{T }}{Q_{0}}\left( \frac{%
p_{T }}{W}\right) ^{\lambda /2} , 
\label{taudef}
\end{equation}
where $p_{T}$ is given in GeV and $W$ in TeV.

For asymmetric systems, the overlapping region is not spherical but rather one-dimensional (see Fig.~\ref{fig:geometry}).
Therefore, we should have 
\begin{equation}
S_{T} \sim n_{\rm{part}}^{2/3}/Q_{0}^{2} ,
\label{STsmall}
\end{equation}
where $n_{\rm{part}}$ is a number of participants in a small system.

In this paper, we consider only d+Au asymmetric collisions, such that we take $n_{\rm part}=1.62$ following \cite{Adare:2013nff}. 
In order to estimate the length $l$ shown in Fig.~\ref{fig:geometry} we take $l=\sqrt[3]{A}=\sqrt[3]{197}=5.82$, following 
Ref.~\cite{Klein-Bosing:2014uaa}. Effectively, this means that for the minimum bias d+Au collisions, the number of participants 
in the  Gold nucleus is $N_{\rm part}[{\rm Au}]=A^{1/3}$, rather than the one determined in Ref.~\cite{Adare:2013nff}, where 
$N_{\rm part}[{\rm Au}] = 7\pm0.4$. Meanwhile, note that the volume of the cylinder shown in Fig.~\ref{fig:geometry} ({\em i.e.}, 
the total number of participants) is equal to
\[ V_{\rm cyl.}=\sqrt[3]{197}\times 1.62^{2/3}=8.03, \]
being quite close to $N_{\rm part} = 8.6\pm0.4$ in minimum bias d+Au, as shown in \cite{Adare:2013nff}.

In view of the above-mentioned discussion, one can write that
\begin{equation}
\bar{Q}^{2} = Q_{0}^{2}\; {N_{\rm part}}/{n_{\rm{part}}^{2/3}} .
\end{equation}
Hence the saturation scale of a large nucleus probed by a small one is equal to:%
\begin{equation}
Q_{\rm{sat}}[{\rm large}] = Q_{0} \left( \frac{N_{\rm{part}}}{n_{%
\rm{part}}^{2/3}}\right) ^{1/2}\left( \frac{p_{T }}{W}\right)
^{-\lambda /2} .
\end{equation}%
For a small nucleus nothing changes with respect to the symmetric collisions, and%
\begin{equation}
Q_{\rm{sat}}\left[ \rm{small}\right] = Q_{0}\,n_{\rm{part}}^{1/6}
\left( \frac{p_{T }}{W}\right) ^{-\lambda /2} .
\end{equation}%
It has been argued in Ref.~\cite{Dumitru:2001ux} that in an asymmetric case the
{\rm effective} saturation scale takes the following form:
\begin{eqnarray}
Q_{\rm sat}^{\rm eff}&=&\sqrt{Q_{\rm{sat}}[{\rm large}]\,Q_{\rm{sat}}[{\rm{small}}]} 
\nonumber \\%
&=& \left( \frac{N_{\rm{part}}^3}{n_{\rm{part}} }\right)^{1/12}Q_{0}\left( \frac{p_{T }}{W}%
\right) ^{-\lambda /2} .
\label{Qsat_asym}
\end{eqnarray}

In order to test the geometrical assumptions described hitherto, we will release the relation 
$S_{T} \sim N_{\rm part}^{2/3} $ in what follows, treating the power of $N_{\rm part}$ as a 
free parameter $\delta$:
\begin{equation}
S_{T} \sim N_{\rm part}^{\delta} ,
\label{STdel}
\end{equation}
which leads to the scaling variables
\begin{eqnarray}
\tau & = &\frac{1}{N_{\rm{part}}^{\delta/4}}\frac{p_{T }}{Q_{0}}\left( \frac{%
p_{T }}{W}\right) ^{\lambda /2} ,
\label{large} 
\end{eqnarray}
and
\begin{eqnarray}
\tau & = &\frac{1}{
\left( N_{\rm{part}}^{3}/n_{\rm{part}}\right) ^{\delta/8}}%
\frac{p_{T }}{Q_{0}}\left( \frac{p_{T }}{W}\right) ^{\lambda /2} ,
\label{tasym}
\end{eqnarray}%
for large-large and small-large collision systems, respectively. For p+p collisions there is no $N_{\rm part}$ dependence.
In this way we are able to test two distinct aspects of GS: the energy dependence by varying parameter $\lambda$,
and dependence on geometry by varying $\delta$. Note that in the limit when the small nucleus size increases, 
$n_{\mathrm{part}}\rightarrow N_{\rm{part}}$, the scaling variable (\ref{tasym}) reduces to (\ref{large}).

The exact coefficient linking  $S_{T}$ and the number of participants  may be of importance for small nuclei, {\em i.e.}, 
for $n_{\rm part}$ (\ref{STsmall}). We expect it to vary at most between 1 and $2^{2/3}$. 
It is therefore of no numerical importance as far as the scaling variable $\tau$ is concerned, due to the power-like 
suppression with $\delta/8$ in (\ref{tasym}). As a matter of fact, because of rather small exponent in the definition 
of $\tau$ in (\ref{tasym}), rescaling $n_{\rm part}$ by a factor of 2 has rather negligible effect: $2^{1/12}=1.06$. 

\begin{table*}[h!]
\centering
\begin{tabular}{cccrrcl}
\\ \hline
$W$ [GeV] & System & \multicolumn{2}{c}{Centrality} & $N_{\rm{part}}$ & Experiment & References
\\ \hline
200 & Au+Au & c1& 0--20 \% & 277.5& PHENIX & \cite{Adare:2008ab} \\ 
&  &c2& 20--40 \% &  135.6 &  &  \\ 
&  &mb& 0--92 \% & 106.3 &  &  \\ \hline
200 & Au+Au & c1& 0--20 \% & 277.5 & PHENIX & \cite{Adare:2014fwh} \\ 
&  & c2& 20--40 \% &   135.6 &  &  \cite{Afanasiev:2012dg}\\ 
&  &  c3& 40--60 \% &56.0 &  &  \\ 
&  & c4 &60--92 \% & 12.5 &  &  \\ \hline
62.4& Au+Au   &mb &0--86 \%  & 114.5 &PHENIX  & \cite{Adare:2018wgc} \\ \hline
200& Cu+Cu   &c1 &0-40 \%  &66.4  &PHENIX  & \cite{Adare:2018jsz}\\
&  & mb &0--94 \% & 34.6 &  & \\ \hline
2760& Pb+Pb &c1  &0--20 \%  &308.0  &ALICE  &\cite{Adam:2015lda} \\
&  &c2&20--40 \%  &157.0  &  & \\
&  &c3&40--80 \%  &45.7  &  & \\ \hline
200 &  d+Au & &      & $ -$  &PHENIX& \cite{Adare:2012vn} \\
 &  p+p     & &  &  $-$ & \\ \hline
\end{tabular}
\caption{The data sets for direct photon $p_{T}$-spectra used in this paper.
For the small system $N_{\rm part}$ see the text.}
\label{tab:data}
\end{table*}

On the contrary, the effect on $F(\tau)$ in (\ref{multHI}) due to the linear factor of $1/S_{T}$ is numerically important. 
It is instructive to have some intuition concerning the possible relation between $S_{T}$ and $n_{\rm part}$. 
Let us go back again to Fig.~\ref{fig:geometry}, where we schematically show the active transverse geometry in 
d+Au and p+p collisions. In d+Au collisions, the average $N_{\rm part}$ of the deuteron related to the transverse size 
$S_{T}$ in (\ref{multHI}) has been determined to be $1.62$ \cite{Adare:2013nff}, due to a large asymmetry of the 
deuteron. The actual transverse size may be larger due to the corona effect.
Here we shall neglect the corona effect in the d+Au case\footnote{The corona effect  has been taken into account in 
Ref.~\cite{Klein-Bosing:2014uaa}, where the authors used $S_{T}\sim (2\,n_{\rm part})^{2/3}=3.2^{2/3}$.}.
On the other hand, in p+p collisions, due to rather a small probability of ``central" collisions, the transverse size
as shown in Fig.~\ref{fig:geometry}, is larger than the size of one proton, so that in this case we simply assume 
$S_{T}\sim 2^{2/3}$. We discuss the sensitivity of our results to these choices in the next section.

\section{\label{sec:GS} Geometrical scaling of direct photon $p_{T}$-spectra}

In our analysis, we have used direct photon data collected in Table~\ref{tab:data}
\cite{Adare:2008ab,Adare:2014fwh,Afanasiev:2012dg,Adare:2018wgc,Adare:2018jsz,Adam:2015lda,Adare:2012vn}. 
These data sets cover rather large range of centrality classes, different beam energies ranging from 62.4~GeV to~2760 GeV, 
and different colliding systems/species. This variety gives us a unique opportunity to study the two distinct 
features of GS: (i) the centrality dependence encoded in the exponent $\delta$, (ii) the energy dependence encoded
in the exponent $\lambda$, both shown in Eqs.~ (\ref{large}) and (\ref{tasym}). In the first case we compare the data
with different $N_{\rm part}$ but at the same energy; initially carrying it out for the large systems only, and then also for 
d+Au, trying to assess the best value of $\delta$. In the second case we take the data at different energies to study the 
GS dependence on $\lambda$. 

Direct photon $p_{T}$-spectra in Au+Au at 200~GeV have also been measured by the STAR Collaboration at RHIC 
\cite{STAR:2016use} several years ago, but those data are not included in our analysis here\footnote{We would expect 
GS holding for the STAR data as well, in spite of the systematic difference between the direct photon invariant yields 
measured by STAR and PHENIX in Au+Au collisions.}. There are also recent results on direct photon $p_{T}$-spectra 
in p+p at 2760~GeV and 8000~GeV beam energies from the ALICE Collaboration at LHC, where upper bounds are given 
\cite{Acharya:2018dqe}.

\subsection{\label{sec:delta} Centrality scaling}

\begin{figure*}[h!]
\centering
\includegraphics[scale=0.7]{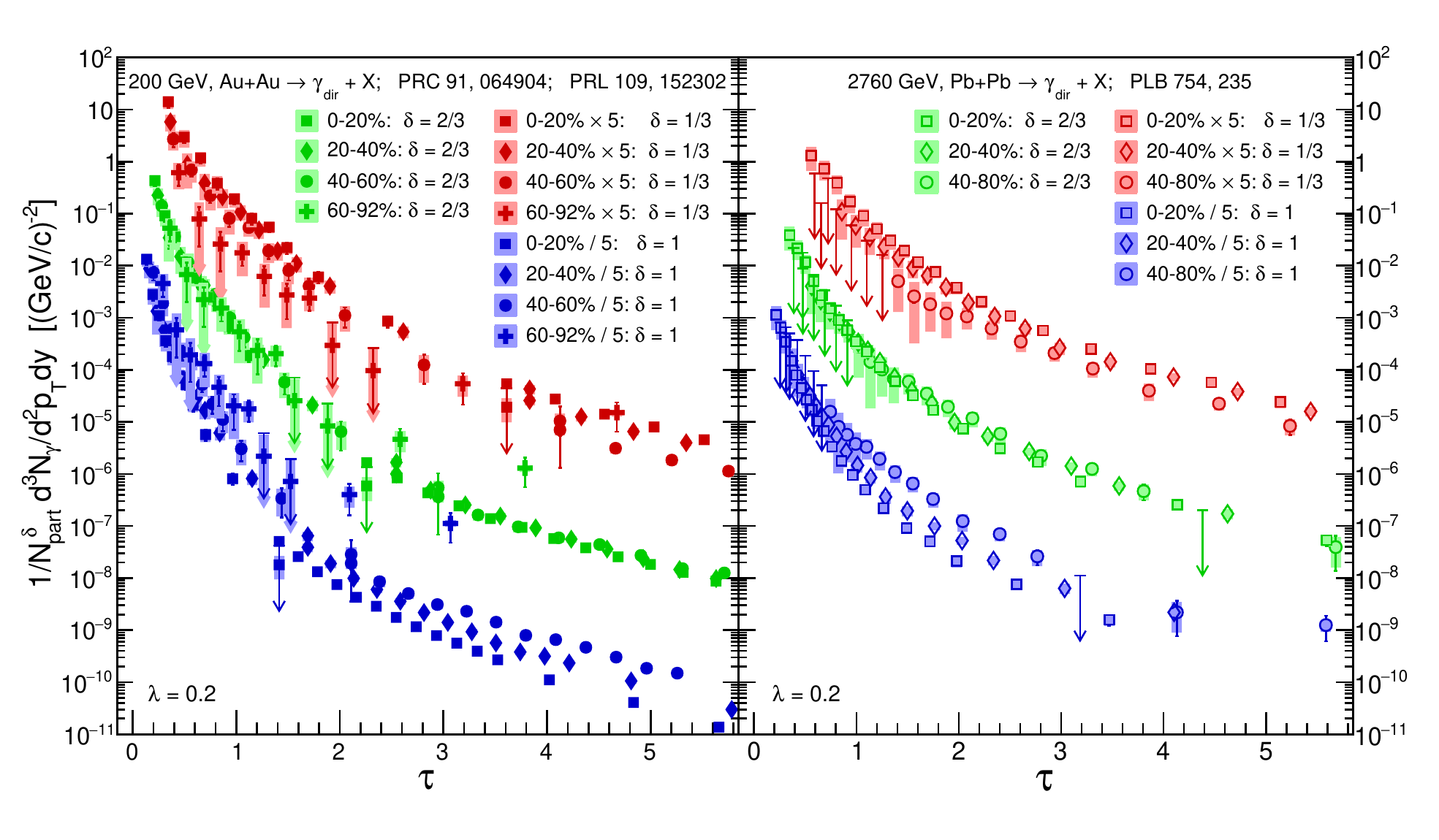}
\caption{(Color online) Direct photon $p_{T}$-spectra scaled according to Eq.~(\ref{multHI}) with $S_{T}$ and $\tau$
given by Eqs.~(\ref{STdel}) and (\ref{large}) respectively, plotted -- from top to bottom -- at $\delta=1/3$ (red points),
2/3 (green points)  and 1 (blue points). The left panel corresponds to plotting of the PHENIX Au+Au data at~200 GeV, 
the right panel to Pb+Pb ALICE data at 2760~GeV (see Table~\ref{tab:data}). The exponent $\lambda=0.2$ does not 
play any role here since we compare data at the same energies.}
\label{fig:delta}
\end{figure*}

In Fig.~\ref{fig:delta}, we plot different centrality-binned  PHENIX 200~GeV Au+Au and 2760~GeV ALICE Pb+Pb direct photon 
invariant yield data, scaled by $1/N_{\rm part}^{\delta}$ in terms of the scaling variable (\ref{large}) for three choices of $\delta$. 
We see that the data at lower $\delta = 1/3$ initially form a wide spread band (upper red points), then come closer to each other 
overlapping at $\delta\approx 2/3$ (middle green points), to disperse again at larger $\delta = 1$ (lower blue points). 
The overlap occurs both for PHENIX and ALICE data at $\delta \approx 2/3$, which is expected on the basis of the GS arguments 
brought up in Sec.~\ref{sec:sat_sca}.

\begin{figure*}[h!]
\centering
\includegraphics[width=7cm]{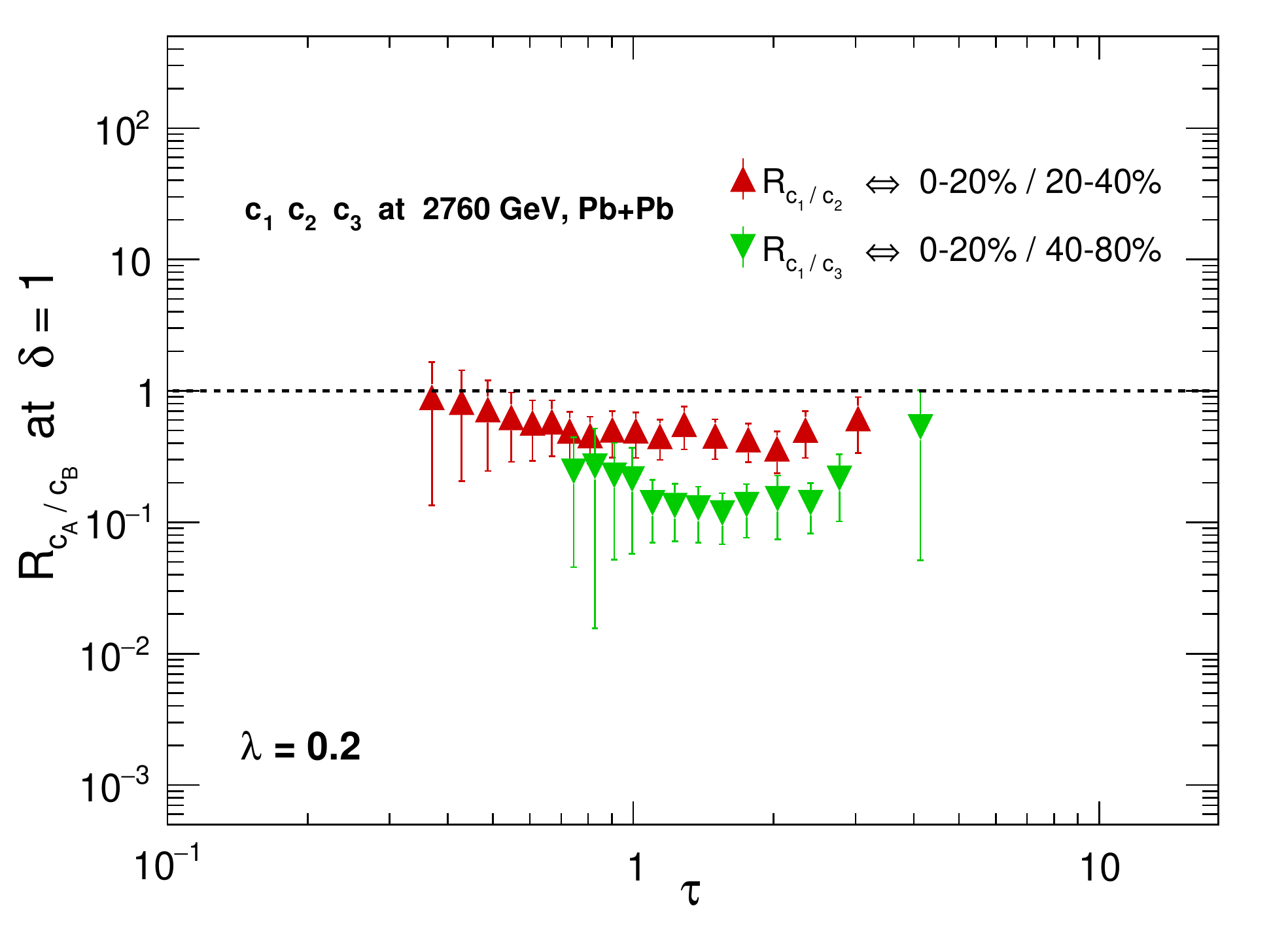}~\includegraphics[width=7cm]{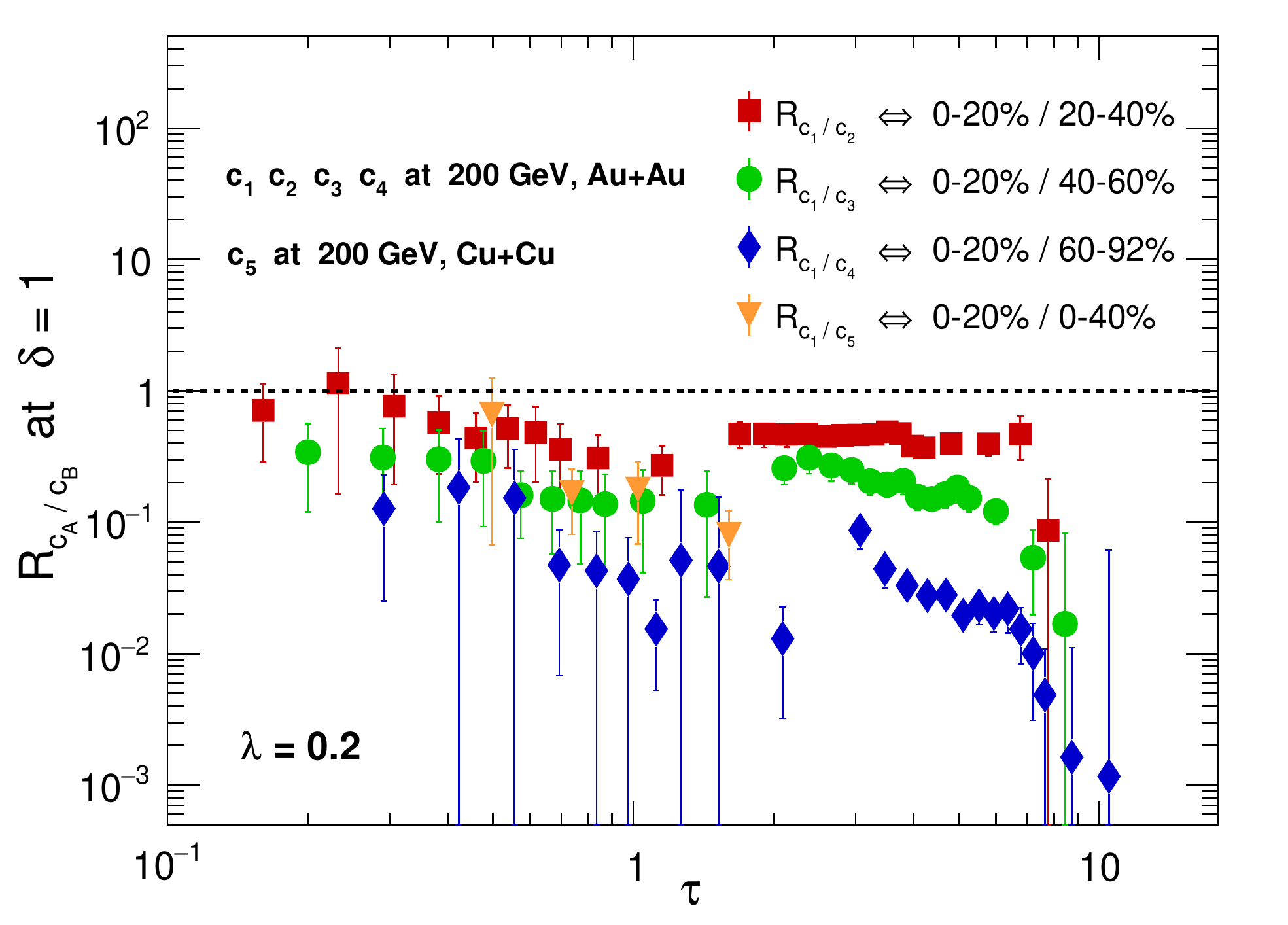}\\
\includegraphics[width=7cm]{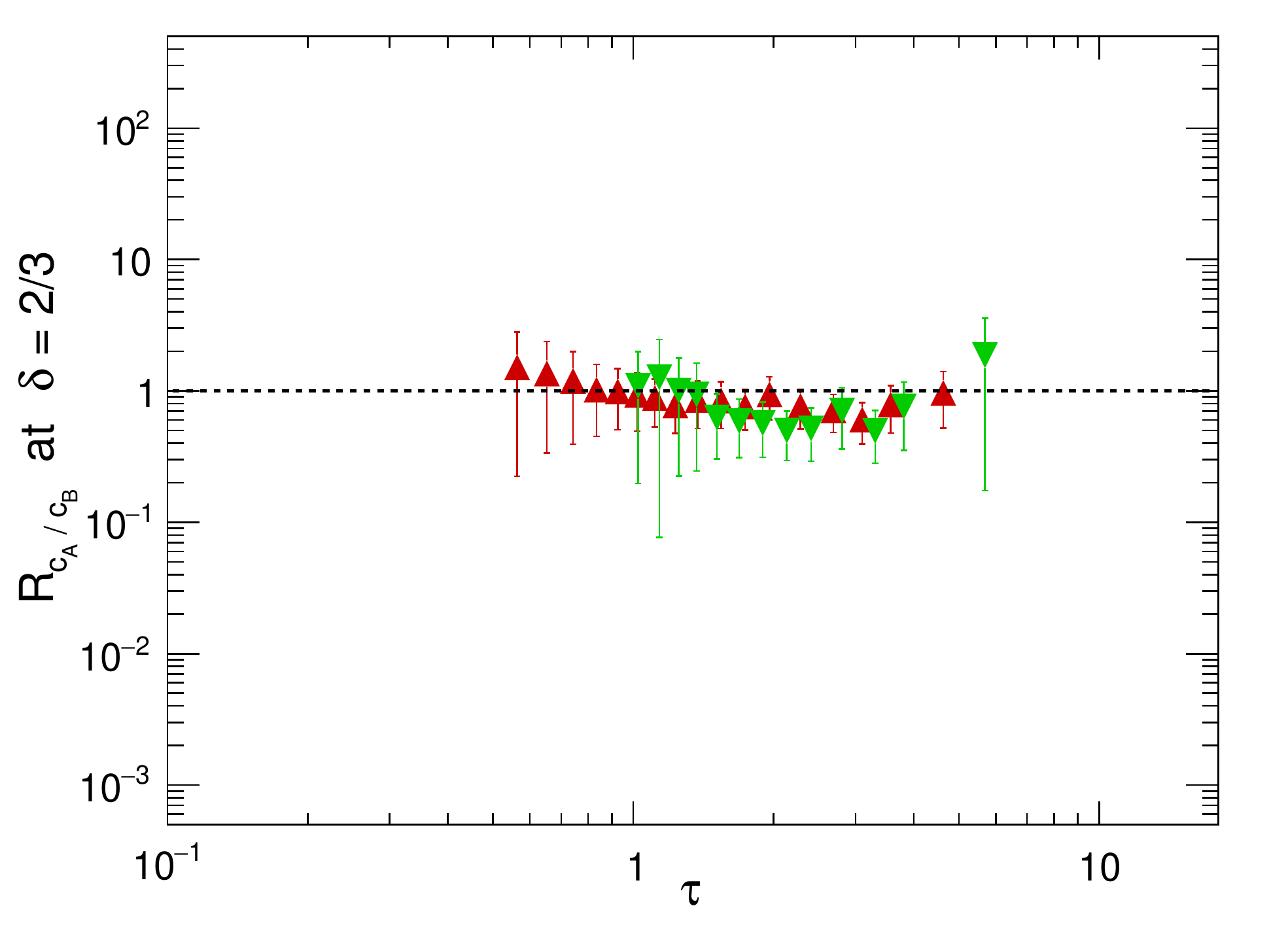}~\includegraphics[width=7cm]{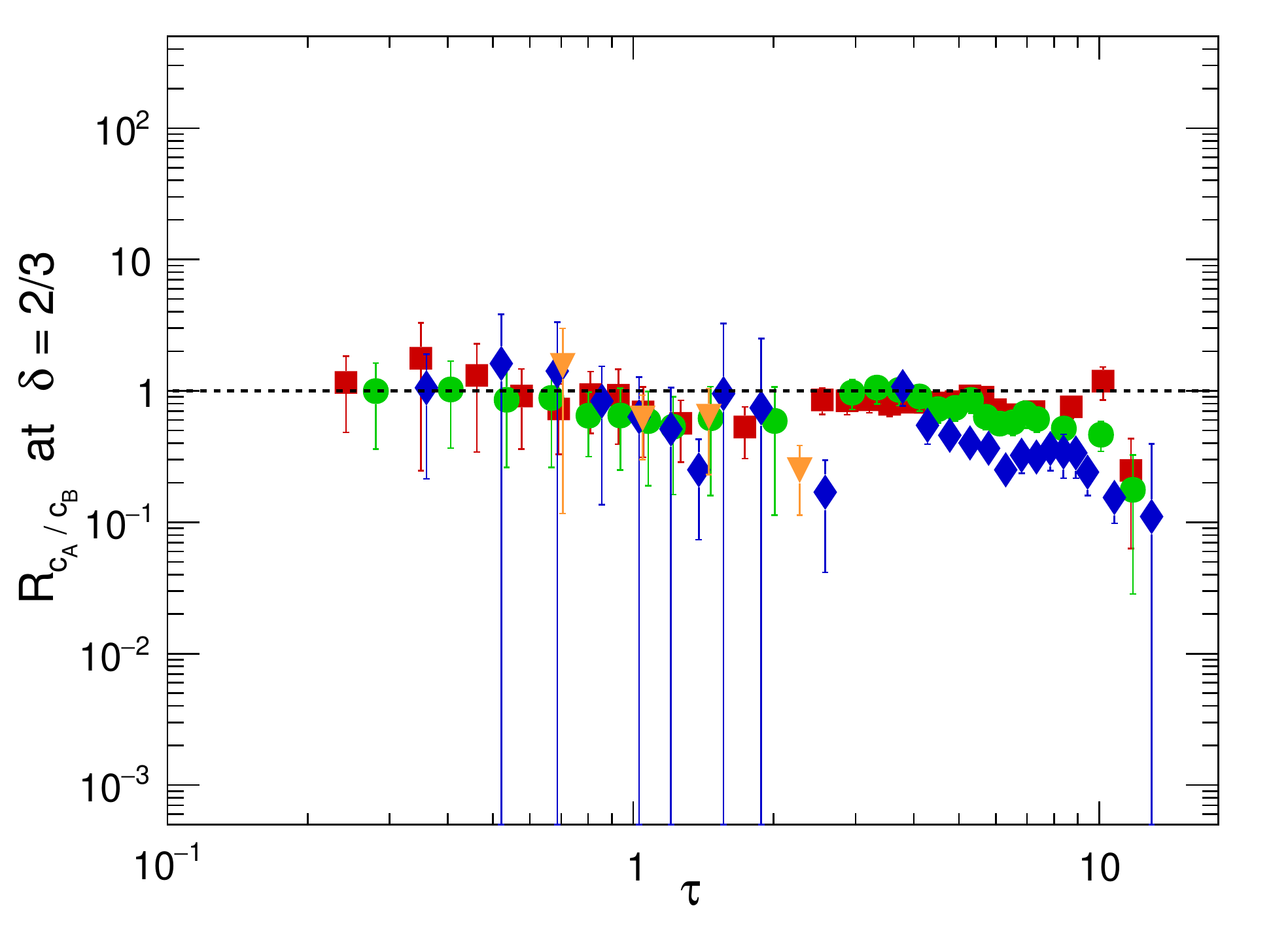}\\
\includegraphics[width=7cm]{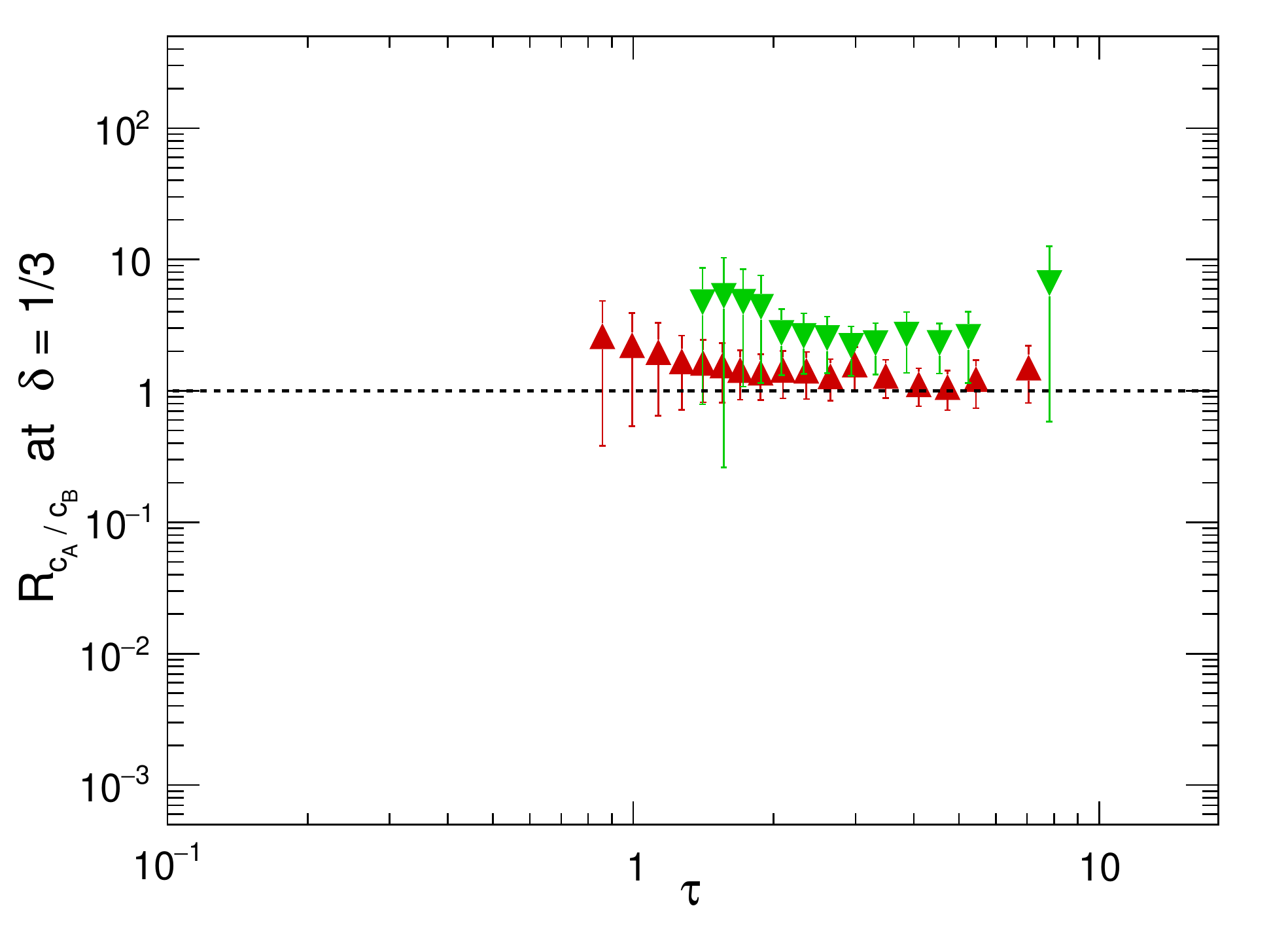}~\includegraphics[width=7cm]{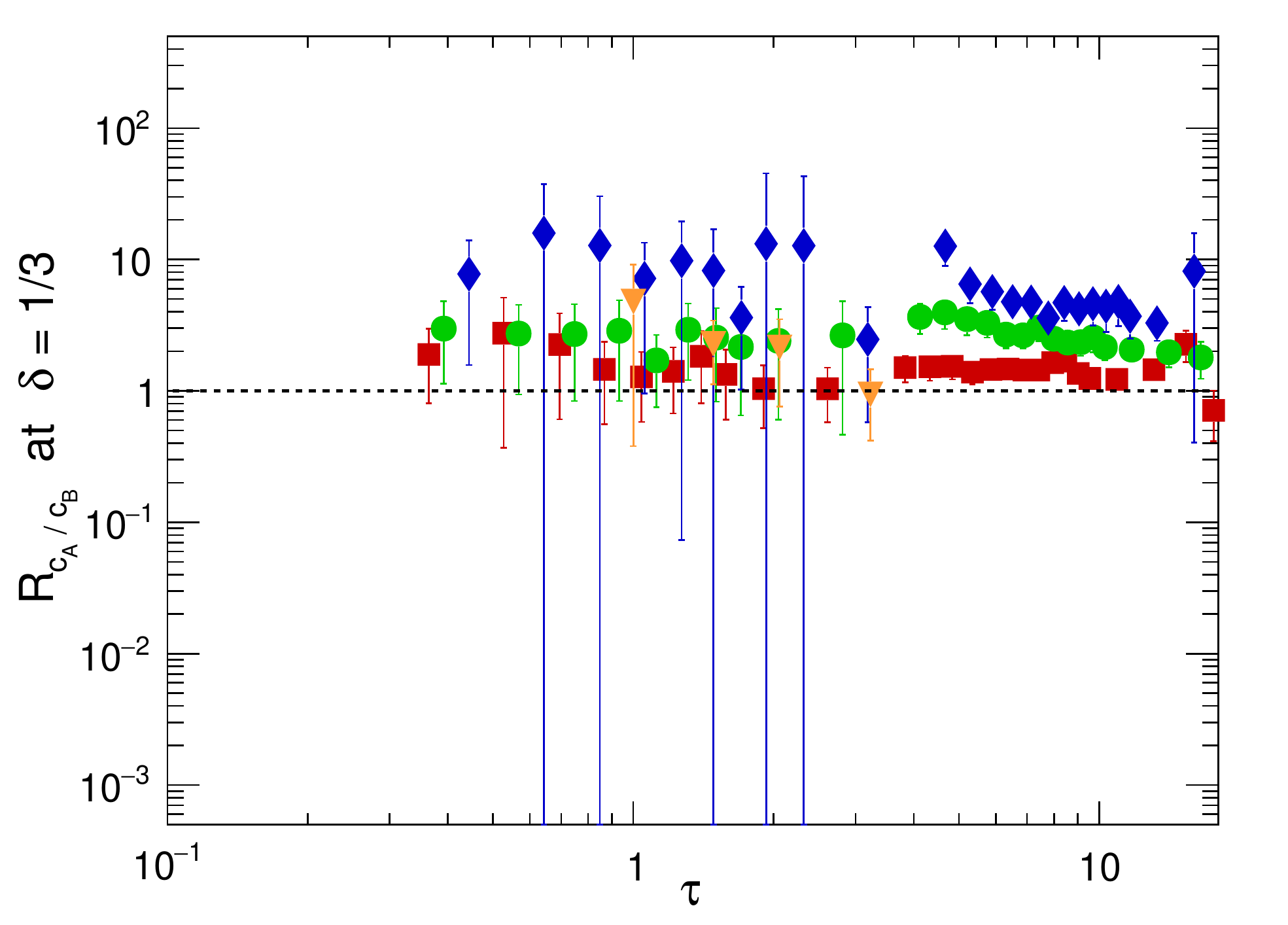}
\caption{(Color online) Another illustration of GS of direct photon $p_{T}$-spectra as ratios of the scaled data in the most central
to the scaled data in the other centrality classes but at the same energies. Left panels: ALICE; the red triangles correspond 
to $R_{c_1/c_2}$ and the green ones to $R_{c_1/c_3}$ for Pb+Pb collisions. Right panel: PHENIX; the red squares correspond 
to $R_{c_1/c_2}$, the green circles to $R_{c_1/c_3}$, and the blue diamonds to $R_{c_1/c_4}$ for Au+Au collisions. The orange 
triangles are for Cu+Cu collisions. The upper plots are obtained at $\delta=1$, middle plots at $\delta=2/3$, and lower plots 
at $\delta=1/3$.}
\label{fig:alph_ratio}
\end{figure*}

In order to see this with a better resolution, we plot ratios (in Fig.~\ref{fig:alph_ratio}) of the most central scaled spectra to those in 
the other centrality classes at different values of $\delta$ (see also Appendix). These ratios should all be equal to unity if GS is present, 
and in fact this happens to a good accuracy at $\delta \approx 2/3$. 


\subsection{\label{sec:deltasmall} Centrality scaling for small systems}

After there are new direct photon invariant yield data on small systems, for example, in p+p, p+Au, d+Au, ${}^{3}$He+Au 
collisions at RHIC, as well as in p+p, p+Pb collisions at LHC, then our analysis of direct photon GS in small systems will 
certainly be more complete. For now we have at our disposal only two data sets from PHENIX, both at 200 GeV 
(see Table~\ref{tab:data}). The p+p data do not scale, so the d+Au collisions render a more interesting case. 
Here, following the discussion of Sec.~\ref{sec:sat_sca}, we use the scaling variable $\tau$ given by Eq.~(\ref{tasym}),
and also the transverse size entering the scaling function $F(\tau)$ in Eq.~(\ref{multHI}) that is given by Eq.~(\ref{STsmall}). 
This follows from the fact that in asymmetric collisions we have in fact two-saturation scales, and the effective saturation 
scale is taken as a square root of the two (see Eq.~(\ref{Qsat_asym})). We need to have therefore two different 
number of participants, for which -- as discussed after Eq.~(\ref{STsmall}) -- we take $n_{\rm part}=1.6$ as the number 
of participants in deuteron, and $N_{\rm part}=\sqrt[3]{197} \simeq 5.8$ as the number of participants in the Gold nucleus. 
For comparison, we also use the one-saturation scale formulae (\ref{ST}) and (\ref{large}) for $S_{T}$ and $\tau$, 
respectively, given $N_{\rm part}=8.6$. In both cases, we consider that the exponent $\delta$ is fixed to 2/3.

\begin{figure}[h!]
\centering
\includegraphics[width=8cm]{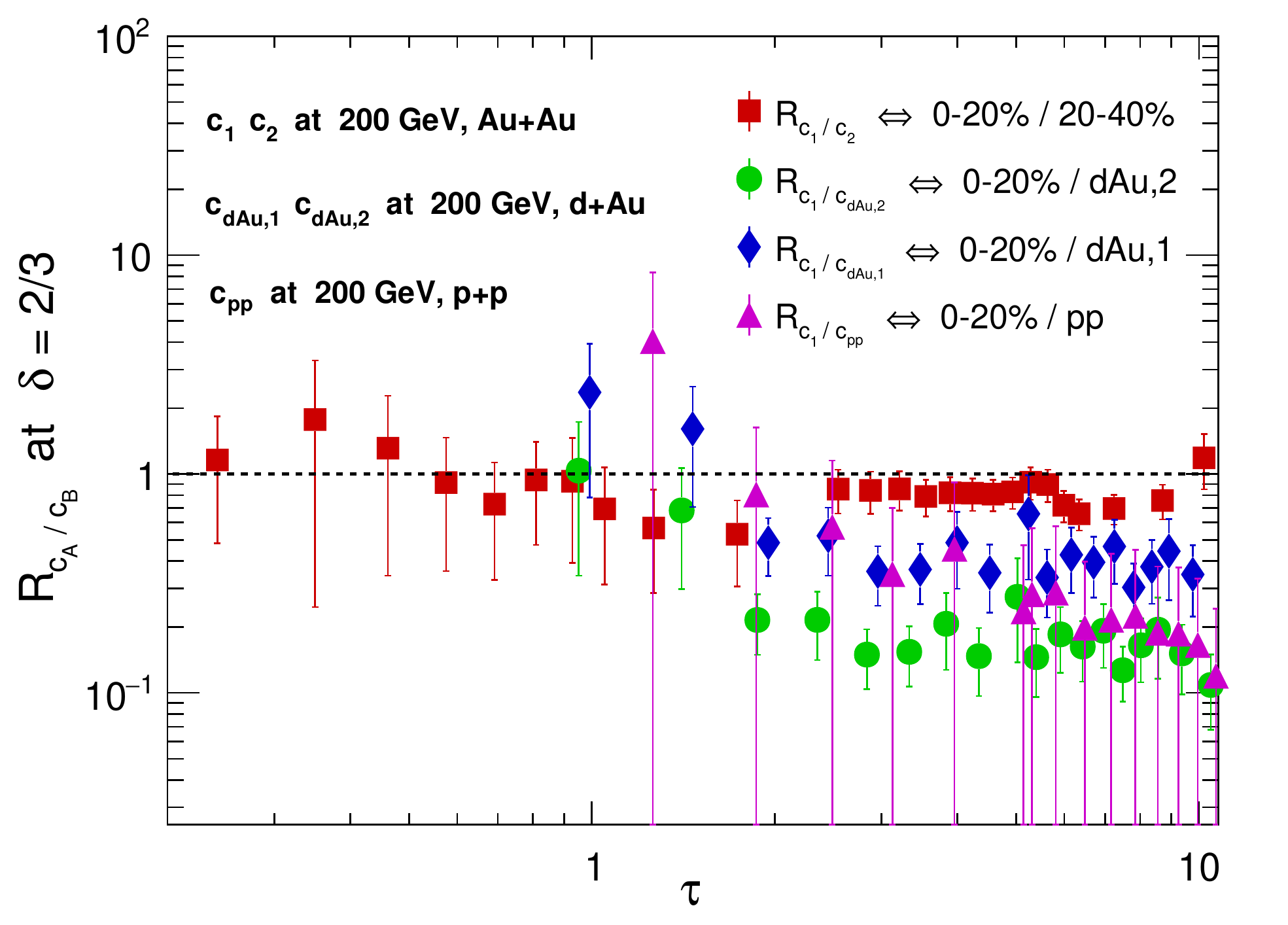}
\caption{(Color online) Similar ratios as in Fig.~\ref{fig:alph_ratio} but for small collision systems. We show the d+Au scaled yields 
(ratio of Au+Au $c_{1}$ over d+Au), with the green circles describing two-saturation scales, and the blue diamonds describing
one-saturation scale (see the text). The red squares correspond to the Au+Au ratio of $c_{1}/c_{2}$ centrality classes, and the 
magenta triangles correspond to Au+Au $c_{1}$ scaled yield divided by the scaled yield in p+p.}
\label{fig:small_system}
\end{figure}

Thus, in Fig.~\ref{fig:small_system} we show the ratios of scaled $p_{T}$-spectra in Au+Au centrality $c_{1}$ to d+A, 
with the two-saturation scales (green circles) and the one-saturation scale (blue diamonds). Also, the ratio of two Au+Au 
centralities $c_{1}/c_{2}$ (red squares) is also given for comparison. Besides, we include the ratio of scaled $p_{T}$-spectra 
in Au+Au $c_{1}$ to p+p  (magenta triangles). At large values of $\tau$, the d+A ratio behaves very much the same as 
the p+p ratio, irrespective whether we use the one- or two- saturation scales formulae. Interestingly enough, a 
difference appears at small $\tau$. The d+Au ratio with the two-saturation scales follows approximately the 
$c_{1}/c_{2}$ ratio of Au+Au collisions. The one-saturation scale formula still shows a sign of 
 GS, albeit within the limits of uncertainties.

We have also investigated the corona effect in d+A collisions discussed in Sec.~\ref{sec:sat_sca}, and we have established 
that it violates GS at small values of $\tau$.

\subsection{\label{sec:lambda} Energy scaling}

\begin{figure*}[h!]
\centering
\includegraphics[width=7cm]{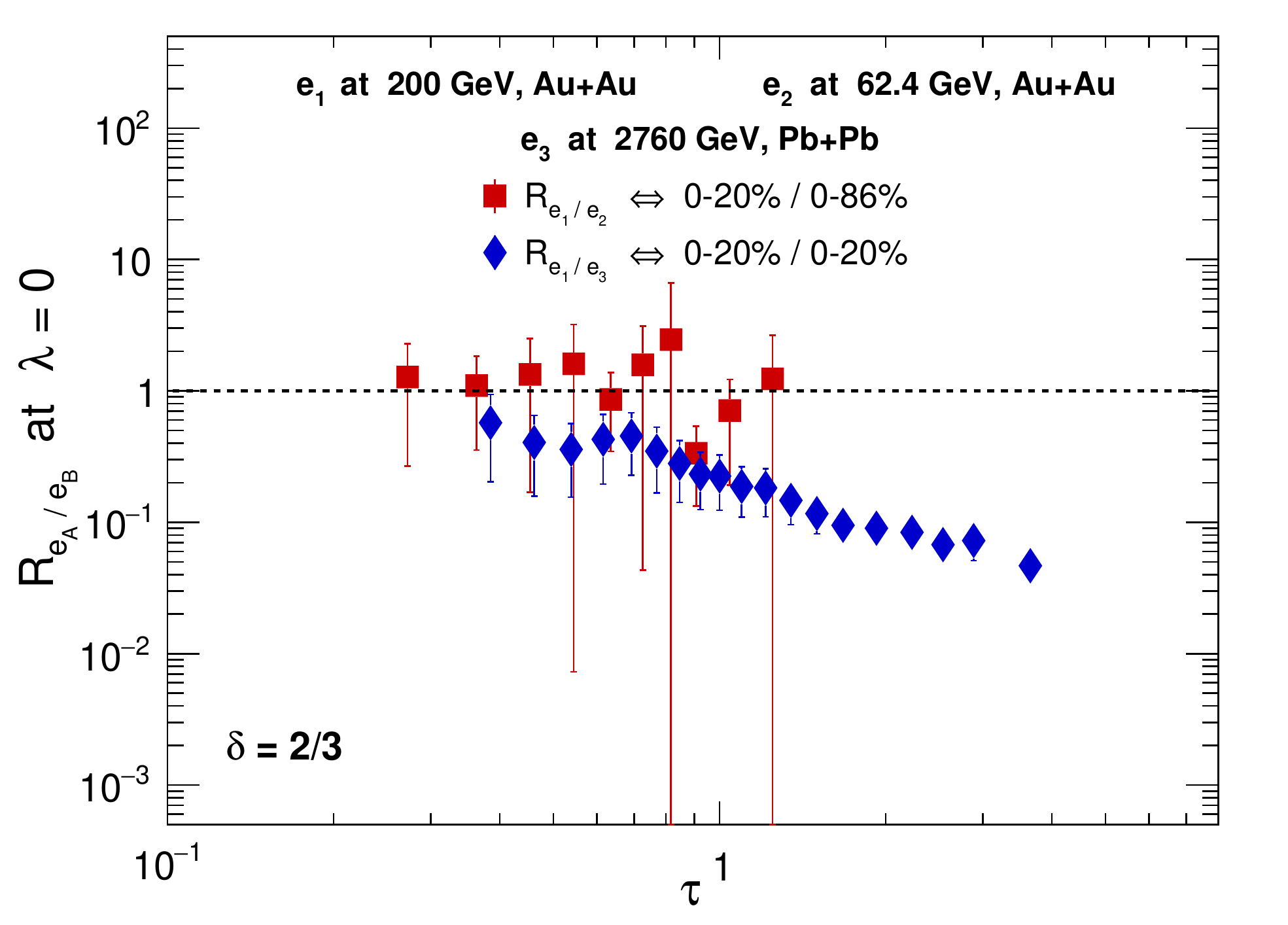}~\includegraphics[width=7cm]{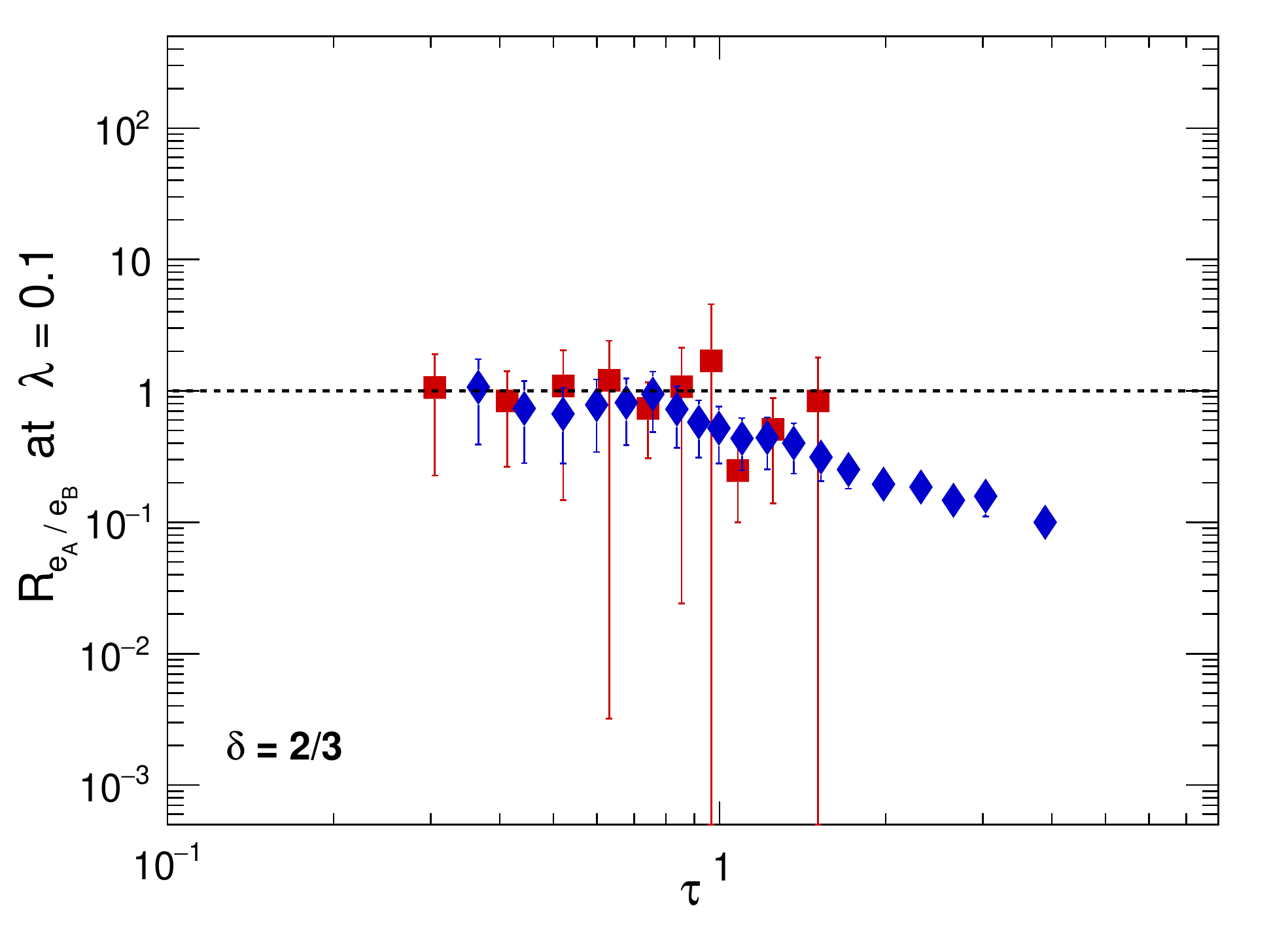}\\
\includegraphics[width=7cm]{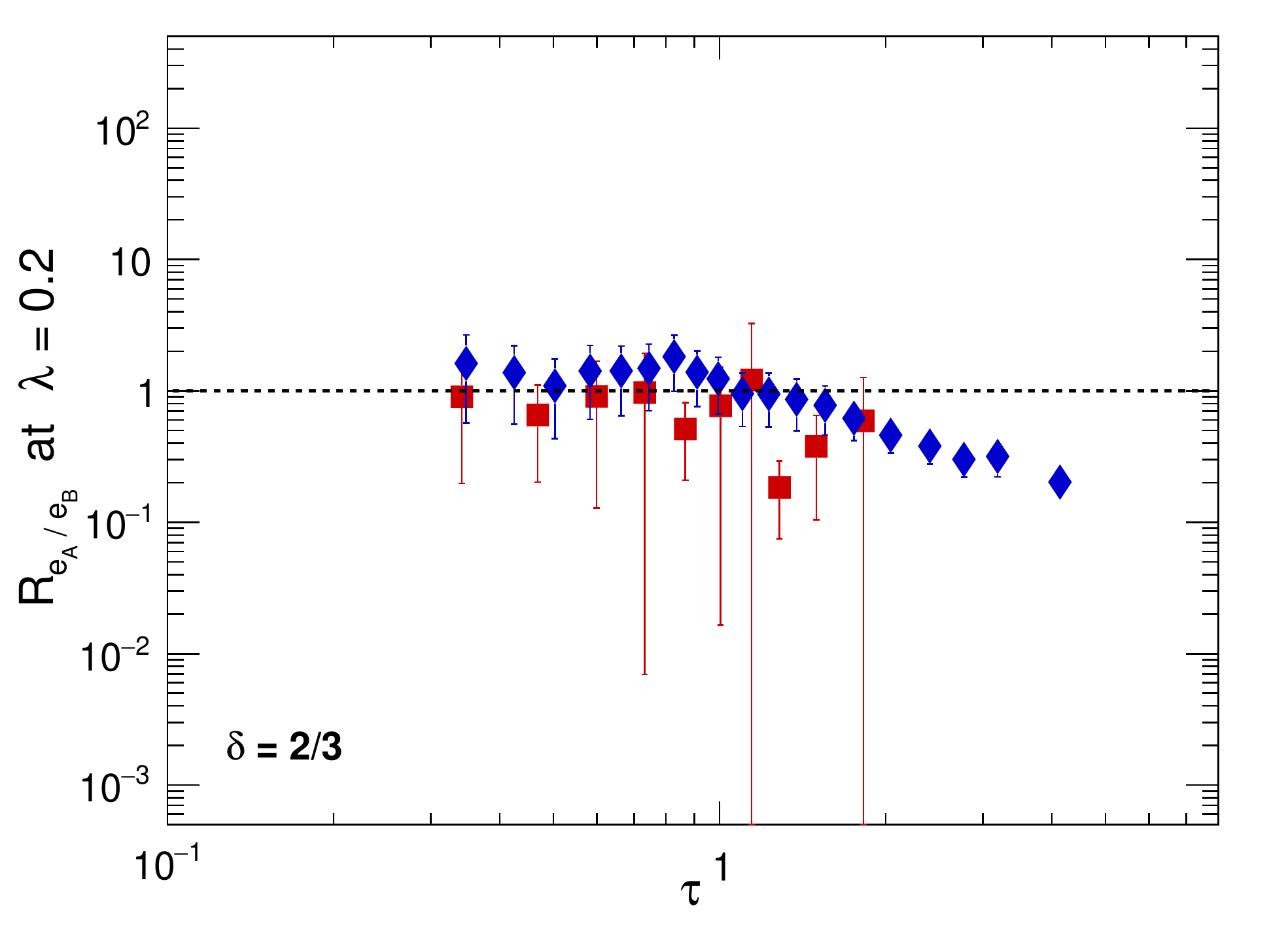}~\includegraphics[width=7cm]{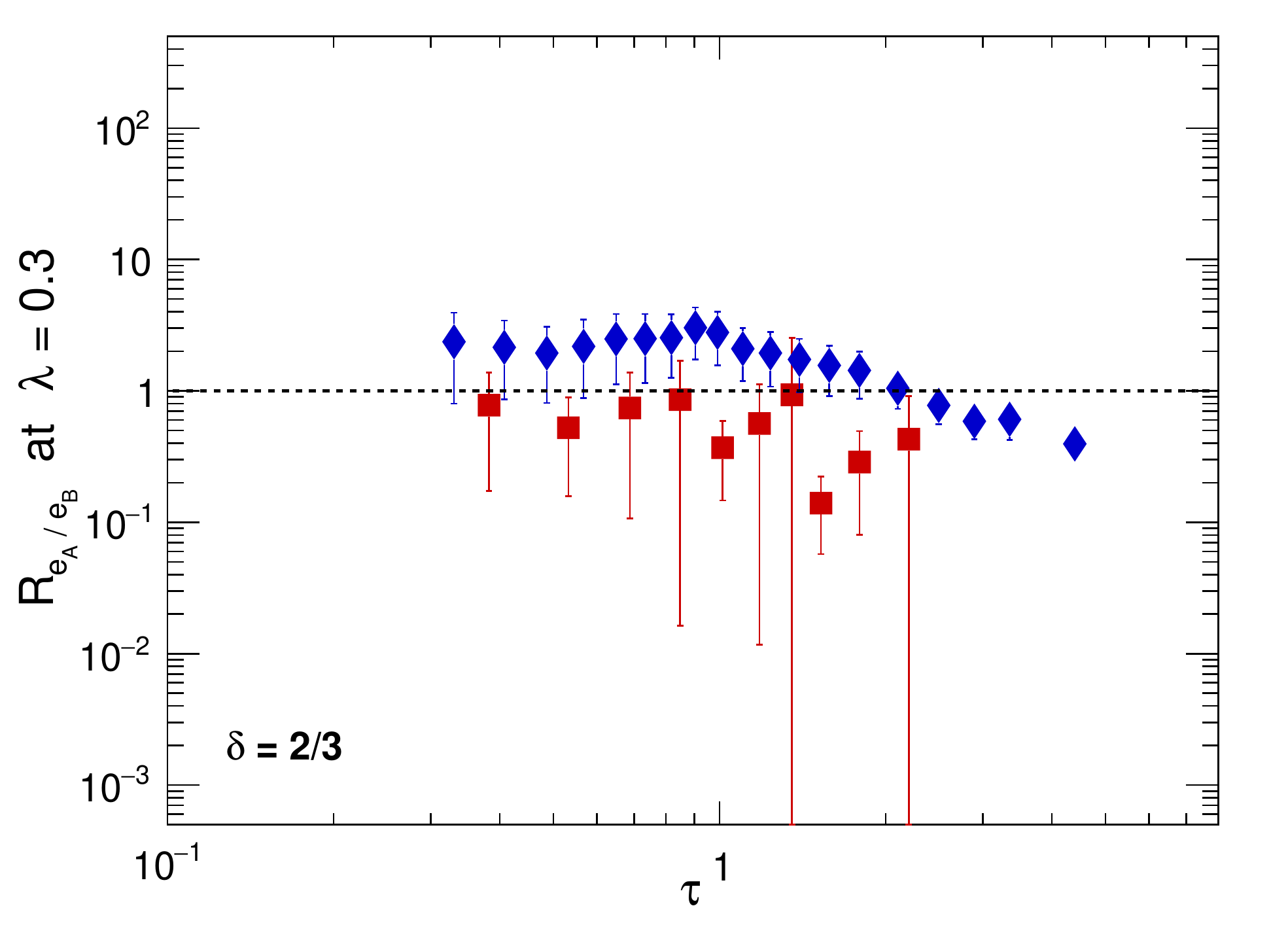}
\caption{(Color online) Illustration of GS of direct photon $p_{T}$-spectra in large collision systems at different energies. 
The figures show ratios of the scaled yields plotted in terms of the scaling variable $\tau$ at three different energies and for four 
different values of $\lambda$ (here $\delta$ is fixed to 2/3): $e_{1}$ at Au+Au at 200~GeV divided by $e_{2}$  at Au+Au at 
62.4~GeV (red squares), and $e_{1}$ at Au+Au at 200~GeV divided by $e_{3}$ at Pb+Pb at 2760~GeV (blue diamonds). 
See text for more details.}
\label{fig:energy_ratio}
\end{figure*}

After having fixed $\delta=2/3$, we can now study the GS dependence on the parameter $\lambda$. To this end, we choose 
one centrality class at different collision energies, namely the most central Au+Au collisions at 200 GeV (designated as $e_1$),
minimum bias Au+Au collisions at 62.4 GeV ($e_2$)\footnote{We have not used the Au+Au data a 39 GeV from \cite{Adare:2018wgc}, 
as they have rather large uncertainties and the collisions energy is not that different from 62.4 GeV.}, and the most central Pb+Pb 
collisions at 2760 GeV ($e_3$). We form two ratios, $e_1/e_2$ and $e_1/e_3$, by plotting them for different values of $\lambda$. 
The result is shown in Fig.~\ref{fig:energy_ratio}. We see that the best scaling is achieved for rather low value of $\lambda$, 
somewhere between 0.1 and 0.2. Certainly, the {\em canonical value}\footnote{Recall that GS in DIS \cite{Stasto:2000er} and in
the {\em cross-sections} in p+p collisions \cite{Praszalowicz:2015dta} is achieved at $\lambda\sim 0.33$.} of $\sim0.33$ is excluded. 
$\lambda=0.2$ is already used in Fig.~\ref{fig:delta} and Fig.~\ref{fig:alph_ratio}. Note that this value is not so far from the 
one used in the {\em multiplicity scaling} of the charged-particle $p_{T}$-spectra in p+p collisions
({\em i.e.} $\lambda=0.22 - 0.24$)~\cite{Praszalowicz:2015dta}.

\section{\label{sec:sumcon} Summary and Conclusions}

In this paper, we have studied the geometrical scaling of direct photon $p_{T}$-spectra (invariant yields) in relativistic heavy
ion and d+Au collisions. We have followed the initial study of Ref.~\cite{Klein-Bosing:2014uaa} with, however, newer data 
and using a method, which does not rely on any specific parametrization of photon yields. If GS is present than all 
$N_{\rm part}^{\delta}$-scaled spectra should be aligned if plotted in terms of the scaling variable $\tau$. If so, their ratios 
should be equal to unity (within experimental uncertainties) for some extended interval of the $\tau$ variable, regardless of 
the centrality class, collision energy, and the colliding nuclei type.

The forms of the scaling variables $\tau$ in (\ref{large}) and (\ref{tasym}), and the scaling function in $F(\tau)$  (\ref{multHI}) 
are dictated by the Color Glass Condensate theory~\cite{McLerran:2010ub}. The fact that we indeed observe GS in HIC, where 
the QGP is produced, means that the information encoded in the initial wave function of colliding systems is transferred
to direct photons. This is a remarkable situation, because GS follows from the scaling properties of the initial wave function of 
gluons, whereas photons are produced from quarks.

GS has two distinct features: it depends on the collision geometry and on the energy behavior of the saturation scale.
We have encoded these two dependencies in the parameter $\delta$ related to the geometry, and the parameter $\lambda$ related 
to the collision energy. While these parameters are in principle fixed to $\delta \simeq 2/3$ and $\lambda \simeq 0.33$ from 
simple geometrical arguments and the DIS studies respectively, we have allowed for their variation in order to asses the best 
values at which GS in HIC is achieved. We have found that the best value for the scaling with centrality is $\delta=2/3$, whereas 
for the scaling with energy we need to take $\lambda$ close to 0.2. The reason why $\lambda$ is substantially lower than its 
expected value $\sim 0.33$ might be based on some scaling violations discussed by us in Ref.~\cite{Khachatryan:2019uqn}, 
and based also on possible other contributions in direct photon production that are not directly related to the initial state physics.

Note that our analysis is not anchored on a detailed $\chi^2$ study. 
One faces here three problems: (i) the accuracy of the alignment of different $p_{T}$-spectra, (ii) how close this alignment is 
to unity (for the ratios of spectra), and (iii) how large is the $\tau$ interval. A good example where the ratios of spectra do align 
but  below unity is the Au+Au $c_{1}/c_{2}$ ratio, and the Au+Au $c_{1}$ to d+Au ratio at small $\tau \sim 1.5$ shown in 
Fig.~\ref{fig:small_system}. We see that both ratios are close to each other, but below one. One should, however, remember 
that in the small $\tau$ region ({\em i.e.}, small $p_T$), mass scales other than the saturation scale come into play 
(particle masses, nonperturbative effects related to $\Lambda_{\rm QCD}$) that spoil dimensional arguments on which 
GS is based. Further studies with possibly upcoming data on small-large collision systems will be of importance to establish 
GS for asymmetric collisions at low-$\tau$ region.

\begin{figure*}[h!]
\centering
\includegraphics[scale=0.42]{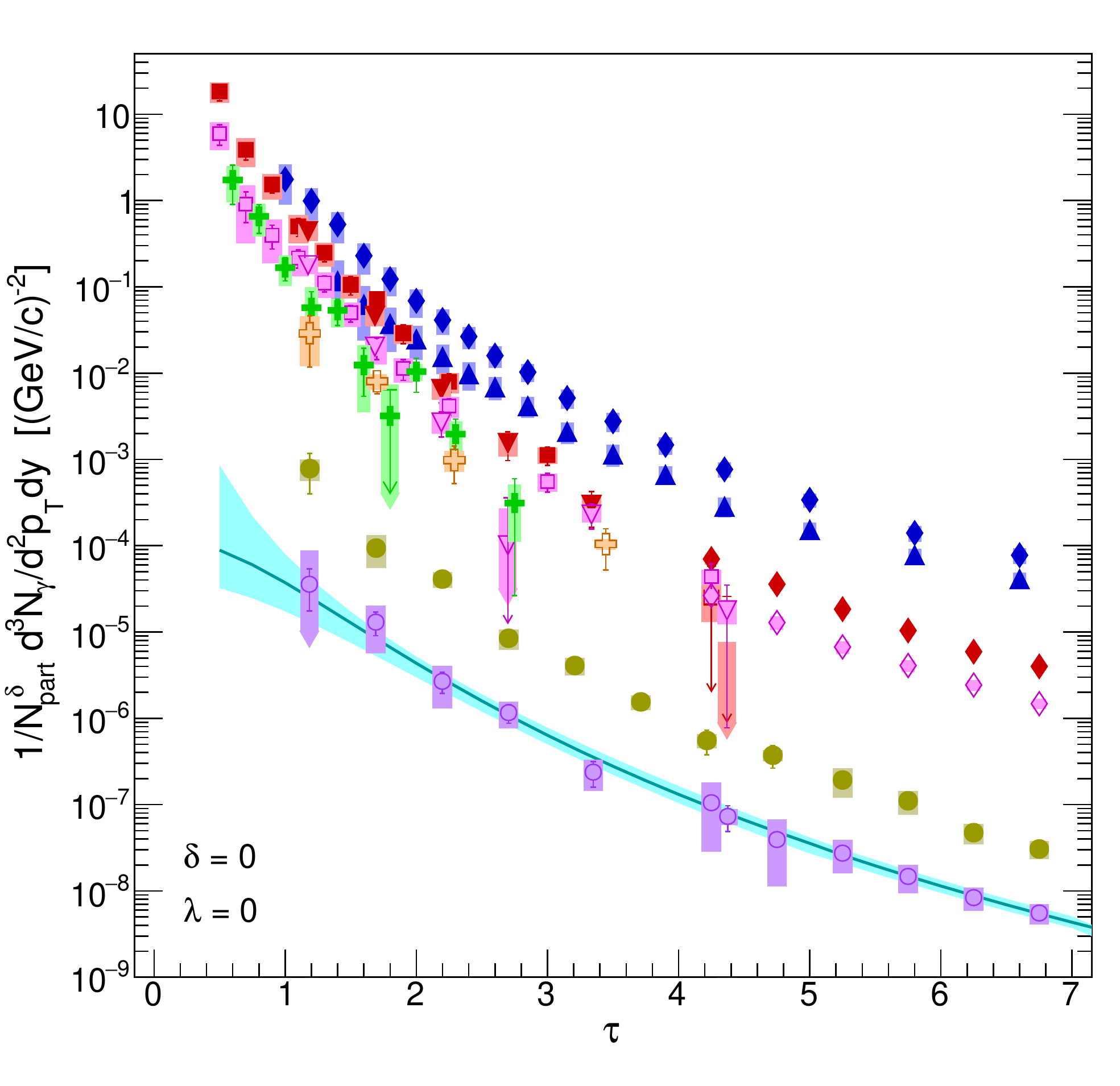}~\includegraphics[scale=0.42]{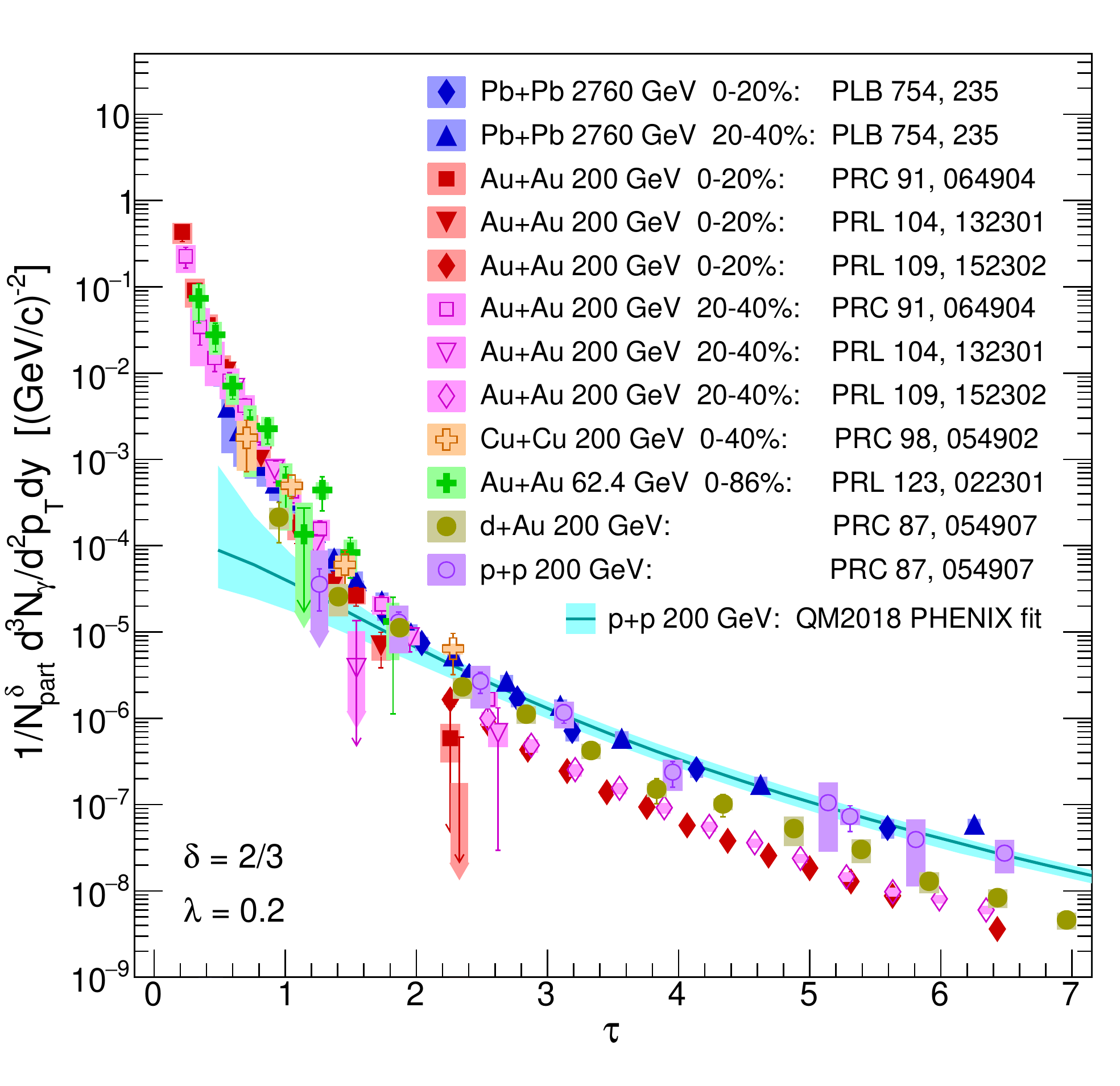}
\caption{(Color online) Direct photon $p_{T}$-spectra of different colliding systems at various energies and centrality classes, all scaled according 
to Eq.~(\ref{multHI}) with $S_{T}$ and $\tau$ given by Eqs.~(\ref{STdel}) and (\ref{large}) respectively. The left figure is obtained at $\delta=0$ 
and $\lambda=0$, {\em i.e.}, shows the original unscaled $p_T$-spectra. The right figure is obtained using $\delta=2/3$ and $\lambda=0.2$. 
The legends show all the references of the data used. The QM2018 PHENIX fit to p+p data is taken from \cite{Khachatryan:2018evz}.}
\label{fig:all}
\end{figure*}

To summarize: we have established GS for direct photon invariant yield data
in HIC and observed a GS sign in d+Au collisions. Our analysis of the p+p data
is rather inconclusive, due to the large uncertainties of the corresponding ratio
shown in Fig.~\ref{fig:small_system}. However, as seen in Fig.~\ref{fig:all} (right panel),
the yield from p+p overall has smaller slope than the yields from HIC. Our final
conclusion is best demonstrated in Fig.~\ref{fig:all}, where we show all data
sets -- non-scaled ones (left panel) and scaled ones (right panel).

\section*{Acknowledgements}

We are grateful to Axel Drees and Larry McLerran for many important and fruitful discussions. 
MP thanks  the Institute for Nuclear Theory at the University of Washington 
for its kind hospitality and stimulating research environment. Research of MP was supported in part by the INT's U.S. 
Department of Energy grant No. DE-FG02- 00ER41132.

\section*{Appendix}

The data points describing direct photon invariant yield in different centrality classes are measured at the
same $p_T$ bins. However, when we represent them in terms of the scaling variables, the values of $\tau$ 
of those points are different at each centrality, as they depend on $N_{\rm part}$. Therefore, if 
we wish to calculate the ratios of the $p_{T}$-spectra in different centrality classes, we need to interpolate 
the data points of at least one centrality to the $\tau$ values, where we have data from another centrality. 
For example, we have selected the PHENIX Au+Au 200 GeV $c_{1} = 0-20\%$ data set as the reference 
spectrum, which we divide by other 
data sets shown in Table~\ref{tab:data}. All this data has been 
published in Refs.~\cite{Adare:2014fwh,Afanasiev:2012dg} for low and high $p_T$ ranges, respectively. 
They have one overlapping point at $p_{T}=4.25$~GeV, for which we take the data point from 
Ref.~\cite{Afanasiev:2012dg}, as it has substantially smaller error than that in \cite{Adare:2014fwh}.

The interpolation is carried out in a simplistic way. For each data point at respective $\tau$ value, we find 
two adjacent data points, such that $\tau_{1} < \tau < \tau_{2}$, and we fit the power law function 
$A/\tau^{n}$ to these two points, both for the yield and for the total uncertainty. We do not take into 
account any additional error coming from this procedure, as for rather densely distributed points such 
an error is much smaller than the experimental uncertainties.



\end{document}